\documentclass{sf2a-conf2015}
\usepackage{graphicx}
\usepackage{hyperref}
\usepackage[]{natbib}  
\usepackage{epstopdf}

\def\BibTeX{{\rm B\kern-.05em{\sc i\kern-.025em b}\kern-.08em
    T\kern-.1667em\lower.7ex\hbox{E}\kern-.125emX}}
\bibpunct{(}{)}{;}{a}{}{,}  

\begin{document}

\TitreGlobal{SF2A 2015}


\title{Clues about the first stars from CEMP-no stars}

\runningtitle{Short title here}

\author{A. Choplin}\address{Department of Astronomy, University of Geneva, Versoix, Switzerland}

\author{G. Meynet$^1$}

\author{A. Maeder$^1$}

\setcounter{page}{237}


\maketitle

\begin{abstract}
The material used to form the CEMP-no stars presents signatures of material processed by the CNO cycle and by He-burning from a previous stellar generation called the source stars. In order to reproduce the relative abundance ratios like for instance C/N or $^{12}$C/$^{13}$C, some mixing between the two burning regions must have occured in the source stars and only the outer layers of the stars, with modest amount coming from the CO core, must have been expelled either through stellar winds or at the time of the (faint) supernova event. With new models at low metallicity including rotational mixing, we shall discuss how the variety of abundances observed for CEMP-no stars can be reproduced.
\end{abstract}



\section{Introduction}

Iron deficient zones in the universe are of particular interest since they were preserved from chemical enrichment, hence delivering clues on the early universe. Carbon enhanced metal poor stars (CEMP) are iron deficient stars with an excess of carbon relatively to the sun. The two common criteria defining a CEMP star are\footnotemark 
\footnotetext{[X/Y] $=\log_{10}$($N_X / N_Y$) - $\log_{10}$($N_{X\odot} / N_{Y\odot}$) with $N_{X,Y}$ the number density of elements X and Y, $\odot$ denoting the abundances in the sun.}
[Fe/H] $< -1$ and [C/Fe] $>0.7$ \citep{aoki07}. s- and r-elements were detected in some of those stars, leading to 4 subclasses : CEMP-s, CEMP-r/s, CEMP-r and CEMP-no stars \citep{beers05}. CEMP-no denote CEMP stars without significant amounts of s- or r- elements. This latter category is of particular interest since it dominates at [Fe/H] $< -3$ \citep{aoki10,norris13}, allowing us to approach the primordial universe even closer. Among the scenarios explored to explain CEMP-no stars, there is the "spinstar" scenario
\citep{meynet06,meynet10,hirschi07,maeder14}, which suggests that CEMP-no formed in a region previously enriched by a material coming from fast rotating, low metallicity, massive stars, experiencing strong mixing, mass loss and eventually a supernova at the end of their lives. For main sequence CEMP-no stars at least, no in situ changes of the surface abundances are expected and observed surface abundances are believed to be the same as at the time of the star formation. Recently, \cite{maeder14} suggested that the variety of CEMP-no abundances can be explained by a material having been processed back and forth by H- and He-burning regions before being ejected by the spinstar. Also shown in \cite{maeder14} is that changing the initial CNO distribution in the spinstar will increase the fit between model and observations in 2D abundance diagrams like $\log (^{12}$C/$^{13}$C)  vs. [C/N].\\ 

We discuss two 32 $M_{\odot}$ models of spinstars computed with the geneva code. Ejecta of the models (wind and supernova) are compared to observed CEMP-no abundances through the 2D abundance diagram [C/N] vs. log($^{12}$C/$^{13}$C). We discuss the impact of taking non-solar initial CNO abundances and emphasize the need for a strong interaction between H- and He-burning shells, in order to synthesize the material needed to build CEMP-no stars.
 
\section{Physical ingredients}

The two 32 $M_{\odot}$ models discussed were computed at $Z=10^{-5}$, with an initial rotation rate\footnotemark 
\footnotetext{$v_{crit}$ is the velocity at which the gravitational acceleration is exactly compensated by the centrifugal force at the equator.} 
of $v/v_{crit} = 0.7$, which corresponds to an initial equatorial velocity of 680 km/s. The evolution is stopped at the end of the neon photodisintegration phase, after the carbon burning. Mass loss prescription is taken according to \cite{vink01} when $T_{eff}>3.95$ and to \cite{jager88} otherwise. We used the recipes of \cite{zahn92} and \cite{maeder97} for horizontal turbulence and shear mixing, respectively. The only difference between the two models is their initial composition : while the first model has a solar-scaled mixture \citep{asplund05}, the second one presents a modified $\alpha$-enhanced mixture ($\alpha$-mod). In the latter mixture [C/N], [O/N] and $^{12}$C/$^{13}$C are put to 2, 1.6 and 30, according to suggestions of \cite{maeder14} for [C/N] and [O/N] and to prediction of galactic chemical evolution models at low metallicity of \cite{chiappini08} for $^{12}$C/$^{13}$C.

\section{Results}

\begin{figure}[ht!]
\centering
\includegraphics[width=0.8\textwidth,clip]{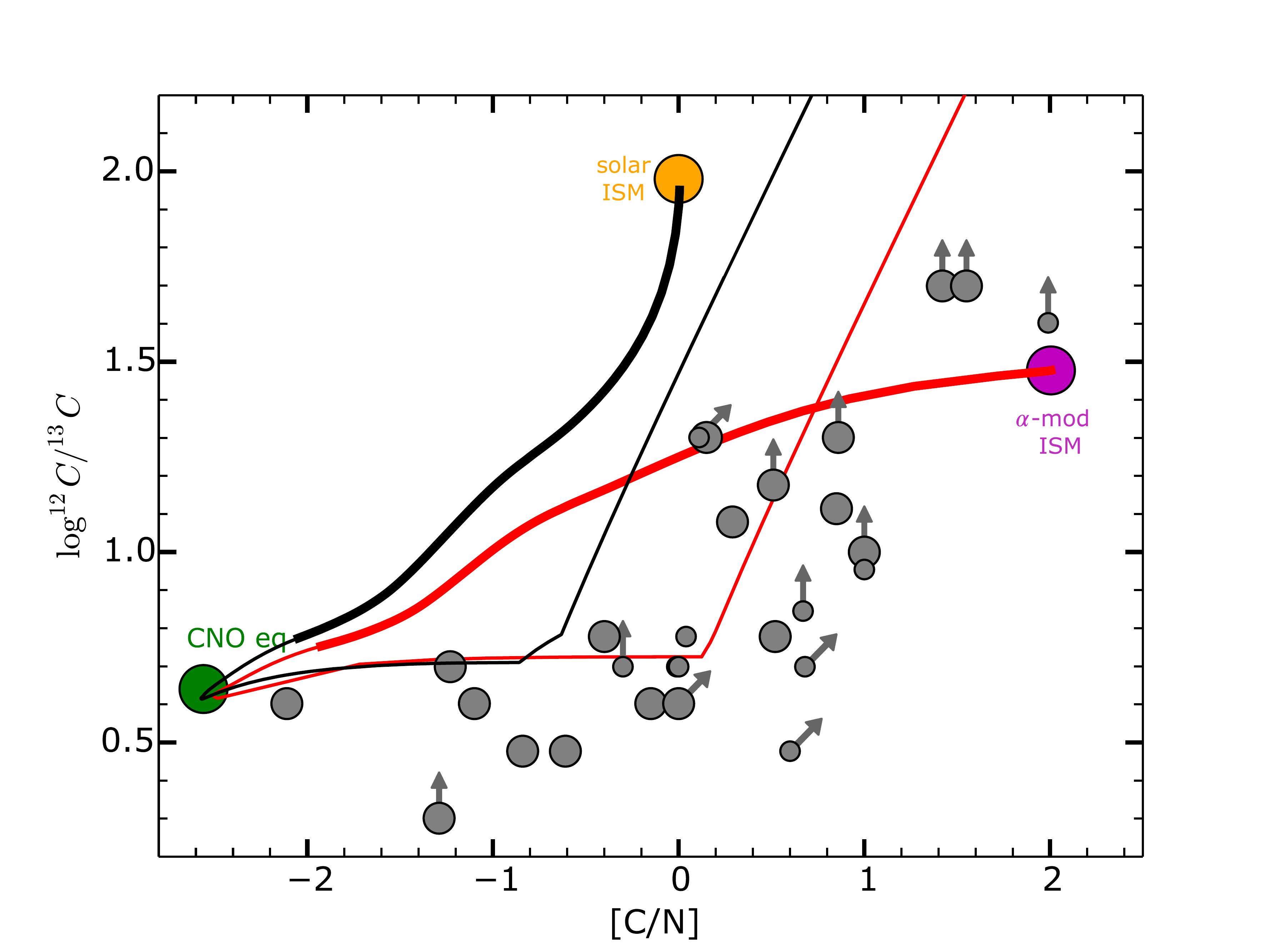}      
\caption{[C/N] vs. $\log(^{12}C/^{13}C)$ diagram. Grey dots are observed CEMP-no stars from \cite{norris13}, \cite{masseron10}, \cite{allen12} and \cite{hansen15} (except for 6 stars, the rest of the sample is the same as \cite{maeder14}, see their table 1 for more details). Small points are MS stars or subgiants while bigger points apply for bright giants.  A vertical arrows is drawn when $^{12}C/^{13}C$ is a lower limit. Oblique arrows indicate a lower limit for $^{12}C/^{13}C$ and an upper limit for Nitrogen at the same time. Yellow and green circles denote values in the sun and when the CNO-cycle is at equilibrium, respectively. Purple circle represent the initial $\alpha$-mod mixture. Thick tracks show integrated abundances ratios in the wind as evolution proceeds and thin lines show the integrated abundances in the ejecta as inner and inner layers of the final stellar structure are added to the wind.}
\label{choplin:fig1}
\end{figure}

Thick tracks in fig.\ref{choplin:fig1} show the integrated abundances in the wind as evolution proceeds for the two models. Comparing the two thick paths, one sees that both are going toward the CNO-equilibrium point : the CNO cycle is at work in such massive stars and transforms $^{12}$C into $^{14}$N and $^{13}$C, leading to lower and lower [C/N] and log($^{12}$C/$^{13}$C) in the wind. Since the initial [C/N] is higher in the $\alpha$-mod mixture, it better fits the observations and cover the whole range of observed [C/N] and almost all the range of $^{12}$C/$^{13}$C. However, it seems that wind cannot provide a material with $-1<$[C/N]$<1$ together with log($^{12}$C/$^{13}$C) $\sim$ 0.5, where most CEMP-no are lying. Moreover, both models lose only $\sim$ 0.5 $M_{\odot}$ through winds, which is probably too low to form a new (even low mass) star. More massive models of spinstars should be modeled since they are likely to eject more mass through winds. A way to get more available mass in the ISM with the present models is to add a supernova to the wind. 

Thin curves in fig.\ref{choplin:fig1} show the integrated abundances in the ejecta as inner and inner layers of the final stellar structure are added to the wind. Although simple, such a supernova simulation allow us to explore all possible mass cuts\footnotemark 
\footnotetext{mass coordinate inside the star delimiting the part which is expelled from the part which is kept into the remant.}.
Even though the red track is going toward higher [C/N], both of those thin curves are similar and can be divided in three parts:

\begin{enumerate}

\item Outer layers ( 20 $< M_{cut} <$ 31 $M_{\odot}$) of the stars are added to the wind. The CNO cycle was at work here, so that the ejecta is more and more enriched in $^{13}$C and $^{14}$N, reducing [C/N] and log($^{12}$C/$^{13}$C) ratios from $\sim -2$ and $\sim 0.75$ to $\sim -2.6$ and $\sim 0.6$ (see fig.\ref{choplin:fig1}).

\item Middle layers are added  (13 $< M_{cut} <$ 20 $M_{\odot}$). $^{12}$C/$^{13}$C is at CNO equilibrium in this region but not [C/N] : there is an excess of $^{12}$C compared to $^{14}$N. This $^{12}$C comes from the He-burning shell which has strongly interacted with the H-shell during the carbon burning phase, leading to [C/N] of $\sim -1$ and 0 for solar and $\alpha$-mod models respectively. Rotational mixing is likely playing an important role to build this special regions in the star, where some He-burning products have moved from the He-shell to the H-shell. The physical process leading to a zone in the star with CNO equilibrium values for $^{12}$C/$^{13}$C but not for [C/N] is likely due to the fact that ($^{12}$C/$^{13}$C)$_{eq}$ is reached quicker that [C/N]$_{eq}$ when the CNO-cycle operates.

\item Inner layers are added ($M_{cut} <$ 13 $M_{\odot}$). The He-shell burning shell is reached, leading to a big rise of both [C/N] and $^{12}$C/$^{13}$C since a He-burning region is  $^{12}$C-rich but $^{13}$C- and $^{14}$N-poor.

\end{enumerate}

As we see, $^{12}$C/$^{13}$C ratio is able to constrain $M_{cut}$ : if too deep layers are expelled, $^{12}$C/$^{13}$C in the ejecta become too high compared to the range of observed value. Only modest amount of He-burning region should be expelled : ejecting 1 $M_{\odot}$ of the He-shell of the model with $\alpha$-mod mixture leads to log($^{12}$C/$^{13}$C) $=1.7$, which correspond to maximum observed values.
 
Considering also the wind ejecta, we see that starting with an $\alpha$-mod mixture in the spinstar improves the fit regarding to a solar mixture : all the range in [C/N] is covered and adding a (faint) supernova to the wind provides a material which corresponds to many observed [C/N] and $^{12}$C/$^{13}$C ratios at the surface of CEMP-no stars. A word of caution, however, regarding the ejecta : while wind, and especially mechanical wind \cite{decressin07}, is expected to stay in the neighbourhood of the star, the supernova ejecta is likely to go further so that the different ejecta should be considered separately. Moreover, some dilution with the ISM can occur, leading to a mixed material made of processed and pristine material. Those points should be taken into account in the future.

One of the other next step is to consider more chemical species (O, Ne, Na, Mg, Al...) in order to increase the level of constraint. Different initial masses must also be investigated, especially to have an idea on which kind of progenitor is preferred to form CEMP-no stars.

\section{Conclusions}

We discussed the possibility of building observed CEMP-no stars with ejecta of spinstars. The two models presented differed only by their initial CNO distribution. The model with a non-solar initial CNO distribution improves the fit in two ways : first the initial [C/N] is taken higher than solar, so that the whole range of observed [C/N] is covered with wind ejecta.
Second, CEMP-no with $-1 <$ [C/N] $< 1$ together with log$(^{12}$C/$^{13}$C) $\sim 0.7$ are well covered by the $\alpha$-mod model if adding a supernova to the wind. Various mass cuts in the spinstar can explain different observed CEMP-no. The special material, with CNO equilibrium values for $^{12}C$/$^{13}$C but not for [C/N], is made available during carbon burning phase, owing to a strong interaction between H- and He-burning shells. Regardless of the initial CNO distribution, $^{12}$C/$^{13}$C appears to be a great ratio to constrain the mass cut at the time of the supernova : if too deep layers are added to the wind, He-burning region is reached and $^{12}$C/$^{13}$C becomes too high compared to observed ratios.

\bibliographystyle{aa}  
\bibliography{choplin} 

\end{document}